\documentstyle[12pt,lnfprep,epsfig,wrapfig]{article}
\newcommand{\Header}{
  \begin{tabular}{rl}
  \hspace{-.4cm}\includegraphics{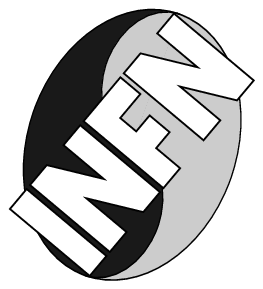} &
    \renewcommand{\arraystretch}{0.5}
    \begin{tabular}{r}
      {\hspace{1cm}~\LARGE\sffamily LABORATORI~ NAZIONALI~ DI~ FRASCATI}\\
      \\
      {\Large\sffamily SIS-Pubblicazioni}\\
    \end{tabular}
    \renewcommand{\arraystretch}{1}
  \end{tabular}
  \vskip 1cm
  \begin{flushright}
  \renewcommand{\arraystretch}{0.5}
    \begin{tabular}{r}
      {\underline{LNF-99/033 (P)}}\\    
      {\small 19 December 1999} \\      
      \\
    \end{tabular}
  \end{flushright}
  \renewcommand{\arraystretch}{1}
  \vskip 1 cm
  }
\def\D*p{ D^{*+} }
\def\D*z{ D^{*0} }

\def\epem{ e^+e^- }
\def\rarr{ \rightarrow }

\begin{document}
\begin{titlepage}
\title{ 
  \Header
  {\large \bf  THE UPGRADED OUTER EM CALORIMETER \\
  OF FOCUS AT FERMILAB }
}
\author{
 S.~Bianco F.L.~Fabbri 
 M.~Giardoni 
 \\
 L.~Passamonti V.~Russo
 S.~Sarwar A.~Zallo 
 \\
 {\it
 Laboratori Nazionali di  Frascati - via E.~Fermi 40, I-00044 Frascati, Italy
 }
 \\
 S.~Carrillo H.~Mendez
 \\
 {\it
 CINVESTAV-IPN, Dept. of Physics, 07000 M\'exico, DF
 }
 \\
 G.~Gianini
 \\
 {\it
 Dip. di Fisica Nucleare e Teorica dell'Univ. and INFN,
 I-27100 Pavia, Italy 
 }
 \\
 J.~Anjos I.~Bediaga C.~Gobel A.~Laudo 
 J.~Magnin \\
 J.~Miranda I.~Pepe F.~Sim{\~a}o 
 A.~Sanchez A.~Reis
 \\
 {\it
 CBPF - R. Dr.~X.Sigaud 150, BR-22290-180 Rio de Janeiro, RJ
 }
}
\maketitle
\baselineskip=14pt

\begin{abstract}
Operational performance, algorithms, stability and physics results of the
 Outer em calorimeter of FOCUS are overviewed.
\end{abstract}

\vspace*{\stretch{2}}
\begin{flushleft}
  \vskip 2cm
{ PACS:11.30.Er,13.20.Eb;13.20Jf;29.40.Gx;29.40.Vj} 
\end{flushleft}
\begin{center}
Presented by F.L.~Fabbri at the VIII International Conference \\
on Calorimetry in HEP, June 13-19, 1999,  Lisbon (Portugal)
\end{center}
\end{titlepage}
\pagestyle{plain}
\setcounter{page}2
\baselineskip=17pt
\section{Introduction and historical overview}
\begin{figure}[t]
 \begin{center}
  \epsfig{file=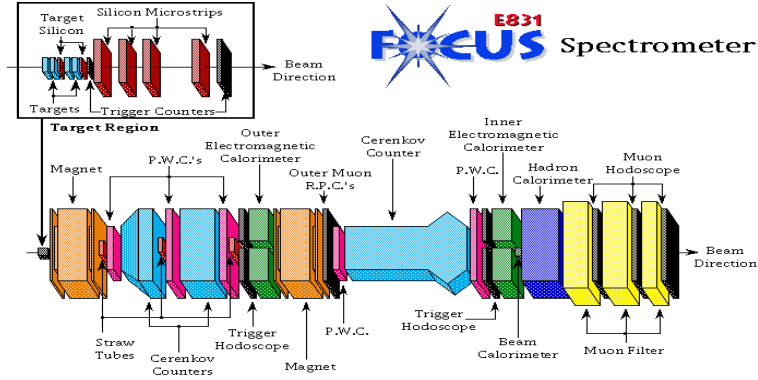,width=10cm,height=5cm}
 \end{center}
  \caption{ The FOCUS (E831) spectrometer at the Fermilab Wide Band Photon
Beam. 
   \label{fig:spec}
 }
\end{figure}

 FOCUS is a heavy-flavor photoproduction experiment located at the 
 Wide Band Photon Area of Fermilab. An upgraded version of its predecessor
 E687\cite{Frabetti:1992au}, 
 FOCUS is composed of 
 (Fig.\ref{fig:spec}) 
 a $\mu$strip silicon detector,
 a large acceptance magnetic 
 spectrometer with MWPC, Cherenkov differential counters, muon detectors,
 hadron calorimetry, and both forward (Inner em), and
 large angle (Outer em) electromagnetic  calorimetry.
 The topics in charm physics that are being investigated include lifetimes,
 semileptonic decays, charm baryons, charm spectroscopy, searches for 
 $D^0-\bar{D^0}$ mixing and for rare and forbidden decays, charmonium
 production and radiative decays, and charm meson and baryon decays with
 neutrals. 
\par 
 The experiment was designed starting in 1981;  in 1985 there was the first
 test  beam  of  the completed spectrometer; in 1987 the first E687
 data-taking   period, 
 interrupted by a    fire. The spectrometer was seriously damaged, but
 quickly repaired. In 1990-1991 the second E687 data taking took place. In
1995 the E687 
 spectrometer became FOCUS and underwent upgrades, with the Outer
 em calorimeter being  equipped with a new plane of square scintillator tiles.
 The 1996-1997 FOCUS data-taking
 period met the goal of collecting ten times the E687 statistics of
 reconstructed charm decays, 
 by fully reconstructing more than ten million charm particle decays.
\section{Physics requirements, geometry acceptances}
\begin{figure}[p]
 \begin{center}
  \epsfig{file=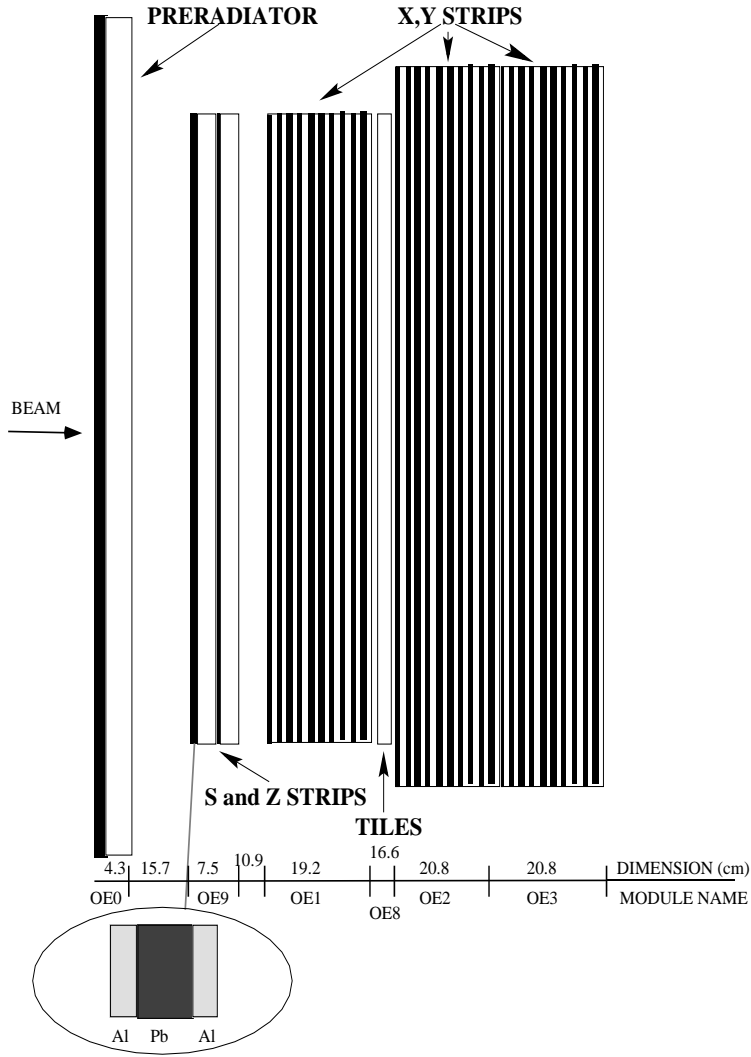,width=10cm,height=15cm}
  \epsfig{file=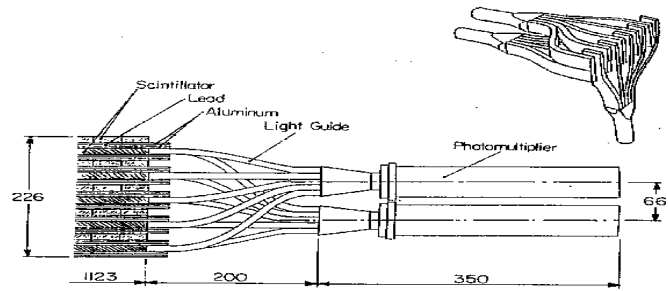,width=4cm,height=9cm,angle=-90}
 \end{center}
  \caption{a) The Outer em calorimeter longitudinal structure showing the
 sandwich of absorber and scintillator planes. b) X and Y
 strip planes interlaced and five-fold integrated to a readout PMT
 (dimensions are millimeters).
   \label{fig:oe}
 }
\end{figure}

 The Outer em  calorimeter (Fig.~\ref{fig:oe}a) is located $900\,{\rm cm}$
 from the target. Its  external  
 dimensions are $(255 \times 205)\,{\rm cm}^2$, with an internal rectangular
 aperture 
 $(51 \times 88)\,{\rm cm}^2$. This corresponds to an angular acceptance for
 photons 
 $(28\leq |\theta_x |\leq 142)\,{\rm mrad}$, $(49\leq |\theta_{y}| \leq
 114)\,{\rm mrad}$. A vertical gap, set at $9\,{\rm cm}$ for the 1996-97
 run, avoids showers from the most abundant background process, i.e.,
 Bethe-Heitler $\epem$ pair production.
 The Outer em is required to reconstruct $\gamma$-initiated showers from 
  charm radiative and $\pi^0$ decays in the energy range $(0.5\leq
  E_\gamma\leq 15)~{\rm GeV}$ and to perform $e/\pi$ identification 
  for charm semielectronic decays in the momentum range $(2.5\leq P \leq
  20)~{\rm  GeV/c}$, thus extending the Cherenkov counter identification,
  which is limited to  $P <6~{\rm  GeV/c}$. Some $\mu/\pi$ identification
 power is  expected to help identify low-momentum muons in charm semimuonic
 decays. 
 Typical geometrical acceptances range from 30\% for electrons and muons in
 charm meson semileptonic decays, and 40-50\% for decays with one or more
 $\pi^0$  in the final state, including the case of shared $\pi^0$ (one
 $\gamma$ in the Inner and one in the Outer em calorimeter).
\section{Mechanical structure}
 The Outer em calorimeter is  made of Pb 
 plates (stiffened with a 6\% Sb by weight) and scintillator 
 layers (POPOP, ${\rm C_{24}H_{12}N_2O_2}$ doped with 8\% 
 naphtalene, and NE-102 were used), for a total of $19\,X_0$ and 
 $1.4\, \lambda_i$ (Tab.\ref{tab:oez}).
 Scintillator layers are made of strips, whose light readout is either 
 individual (OE0, OE9 segments) or five-fold integrated by a light guide 
 to a single PM (all other segments),  for a total
 of 1136 readout channels.
 Each counter in the calorimeter  is individually wrapped in $0.1\, {\rm
 mm}$  Al  foils and black plastic. 
 Horizontal and vertical five-fold counters are interlaced as shown in 
 Fig.\ref{fig:oe}~b).
 The counters are arranged in nine  independent views along Z (Z is the beam
 direction), and four independent quadrants in the (X,Y) plane. A module of
 S-Z strips $(45^o-135^o)$ performs horizontal-vertical matching of
 clusters. 
 The  counters are equipped with ten-stage, EMI-9902KB photomultiplier 
 tubes (PMT) operating at a typical gain 
 of $10^6$ at $1000\,{\rm V}$, with a quantum efficiency of $20\%$ at
 $440\,{\rm nm}$,  
 which were individually tested in 
 order to select only those with good linearity and small sensitivity to rate 
 effect\cite{Bianco:1986gf}.
 The PMT's are powered by LeCroy 1440 and custom-made 
 FRAMM\cite{bologna82}  HV systems, via a high-linearity, anode grounded
 voltage   divider supplying $1.5\, {\rm mA}$ at $1500\, {\rm V}$.
 The PMT signals reach the counting room via $60\,{\rm m}$ long coaxial
 cables, where they 
 are converted by a Lecroy 1881M Fastbus ADC, 
 with a $0.050\,{\rm pC/count}$ conversion.
 The Outer em calorimeter can be displaced both horizontally and vertically
 for  
 calibration and access purposes. 
 \begin{table}
 \caption{Longitudinal segmentation and  counter geometry of the Outer em
 calorimeter. 
   \label{tab:oez}
 }
 \begin{footnotesize}
 \begin{center}
  \begin{tabular}{|l|c|c|c|c|c|} 
  \hline
          & OE0  & OE9S & OE9Z & OE1V & OE1H \\
 \hline\hline
          &      &      &      &      &      \\
 $X_o$ sampled
   &0-1.3 &1.3-1.9&1.9-2.5&2.5-7.3&3.0-7.8   \\
 \hline
 $\lambda_i$ sampled
   &0-.09 &.09-.15&.15-.21&.21-.56&.25-.60   \\
\hline
Sandwich struct.
 & \multicolumn{3}{c|}{AlPbAlSc}
 & \multicolumn{2}{c|}{$5\times$(AlPbAlSc)}   \\
\hline
Pb thick. [cm]
   & 0.650 & \multicolumn{4}{c|}{0.254}        \\
\hline
Al thick. [cm]
   & \multicolumn{1}{c}{}  & \multicolumn{4}{c|}{0.254}        \\
\hline
Scint. type
   & NE102 & \multicolumn{4}{c|}{POPOP}        \\
\hline
Scint. thick. $[cm]$
   & \multicolumn{3}{c|}{3.0} & \multicolumn{2}{c|}{1.0}                \\
\hline
Counter width $[cm]$
   & 3.3   & \multicolumn{2}{c|}{7.0}  & \multicolumn{2}{c|}{3.3}       \\
\hline
Counter hor. 
   & hor   &$45^o$ &$135^o$& vert  &  hor                 \\
\hline
Counters integrated
   &  \multicolumn{3}{c|}{1}       &  \multicolumn{2}{c|}{5}        \\
\hline\hline
 \hline
          & OE8T  & OE2V   & OE2H   & OE3V    & OE3H          \\
 \hline\hline
          &       &        &        &         &               \\
 $X_o$ sampled 
          &7.8-7.9&7.9-12.7&8.4-13.3&13.3-18.2&13.7-18.6      \\
 \hline
 $\lambda_i$ sampled
          &.60-.61&.61-.96 &.65-1.1 &1.1-1.4  &1.14-1.44      \\
\hline
Sandwich struct.
 & \multicolumn{1}{c|}{AlScScAl}
 & \multicolumn{4}{c|}{$5\times$(AlPbAlSc)}   \\
\hline
Pb thick. [cm]
   & -     & \multicolumn{4}{c|}{0.254}        \\
\hline
Al thick. [cm]
   &  \multicolumn{5}{c|}{0.254}        \\
\hline
Scint. type
   & BC404-B & \multicolumn{4}{c|}{POPOP}        \\
\hline
Scint. thick. $[cm]$
   & \multicolumn{1}{c|}{0.5} & \multicolumn{4}{c|}{1.0}                \\
\hline
Counter width $[cm]$
   & 10.0   & \multicolumn{4}{c|}{3.3}       \\
\hline
Counter hor. 
   & sqr   & vert  &  hor  & vert  & hor                \\
\hline
Counters integrated
   &  \multicolumn{1}{c|}{2}       &  \multicolumn{4}{c|}{5}        \\
 \hline
 \end{tabular}
 \end{center}
 \end{footnotesize}
\end{table}
\par
 A scintillator tile array
 module  recovers  showers in the small-angle, high-occupancy region, improves
 horizontal-vertical matching,
 and cleans the $\pi^0$ peak by rejecting fake matches.
 The  module is located at shower max, i.e., between the OE1 and OE2
 modules, and is composed of an 
 array of 100 supertiles, and edge counters to flag laterally
 noncontained showers. Each supertile is made of two  
 $(10\times 10 \times 0.5)\, {\rm  cm}^3$ tiles (BICRON BC404-B
 scintillator), each is 
 equipped with two 20~cm-long,  1-mm-diameter wavelength shifting (WLS)
 optical 
 fibers (Kuraray Y11 multiclad S-type) 20~cm-long, $\alpha$-cut, with all
 four ends thermally spliced to 2~m-long clear
 fibers  following the CDF endplug splicing 
 technique \cite{Apollinari:1992zf}, with a heat-shrinking
 tube to protect the splice.  
 The eight ends of the clear fibers
 of each supertile are optically coupled by means of optical grease to a
 EMI-9902KB PMT. 
\begin{wrapfigure}[20]{r}{5cm}
 \begin{center}
  \epsfig{file=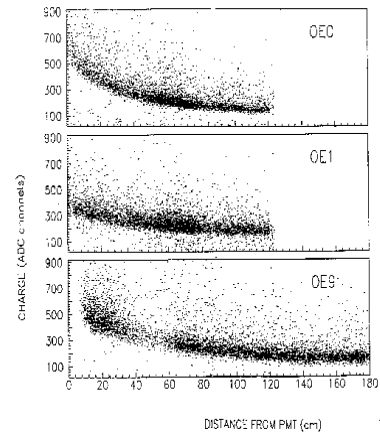,width=5cm,height=6cm}
 \end{center}
 \caption{ Light attenuation curves for three different sized strip counters.
    \label{fig:mips}
 }
\end{wrapfigure}
 Each tile is wrapped in white Teflon tape, and the side is
 painted with white reflective paint by BICRON.
 The tile array is enclosed
 in a light-tight Al case. The light transmission efficiency of the thermal
 splice was measured on relevant samples during the splicing
 process  and shown to be typically 94\%.
\begin{figure}[t]
 \begin{center}
  \vspace{5cm}
  \includegraphics{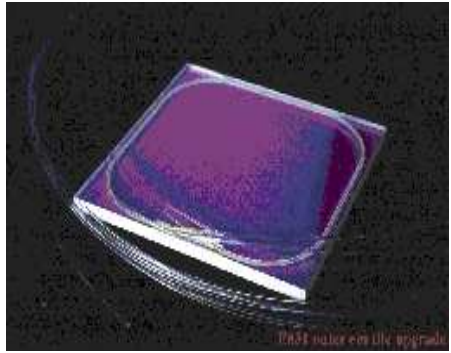}
  \includegraphics{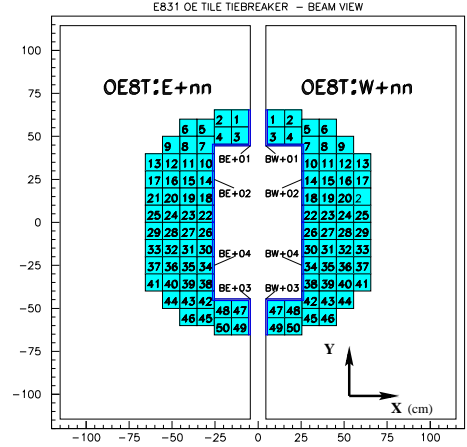}
 \end{center}
 \caption{ left) A complete tile counter before wrapping; 
 right) layout of tile counter array, and edge counters.
   \label{fig:tiles}
 }
\end{figure}
\section{Equalization with mips }
 Beam halo muons were used to determine the counter geometry,
 for an approximate PMT gain balancing ($\pm
 10\%$), to determine the light attenuation curves inside strips
 (Fig.\ref{fig:mips}), and the equalization constants. 
  The light output was measured in the laboratory with cosmic rays
 to be    30  photoelectrons/mip for a supertile,
 and 100 photoelectrons/mip 
 for a five-fold counter at $20\,{\rm cm}$ from PMT. The relative width fwhm
 of the energy distribution for a mip is 40\% for a supertile, and 30\% for
 a five-fold counter.
\begin{figure}[t]
 \begin{center}
  \epsfig{file=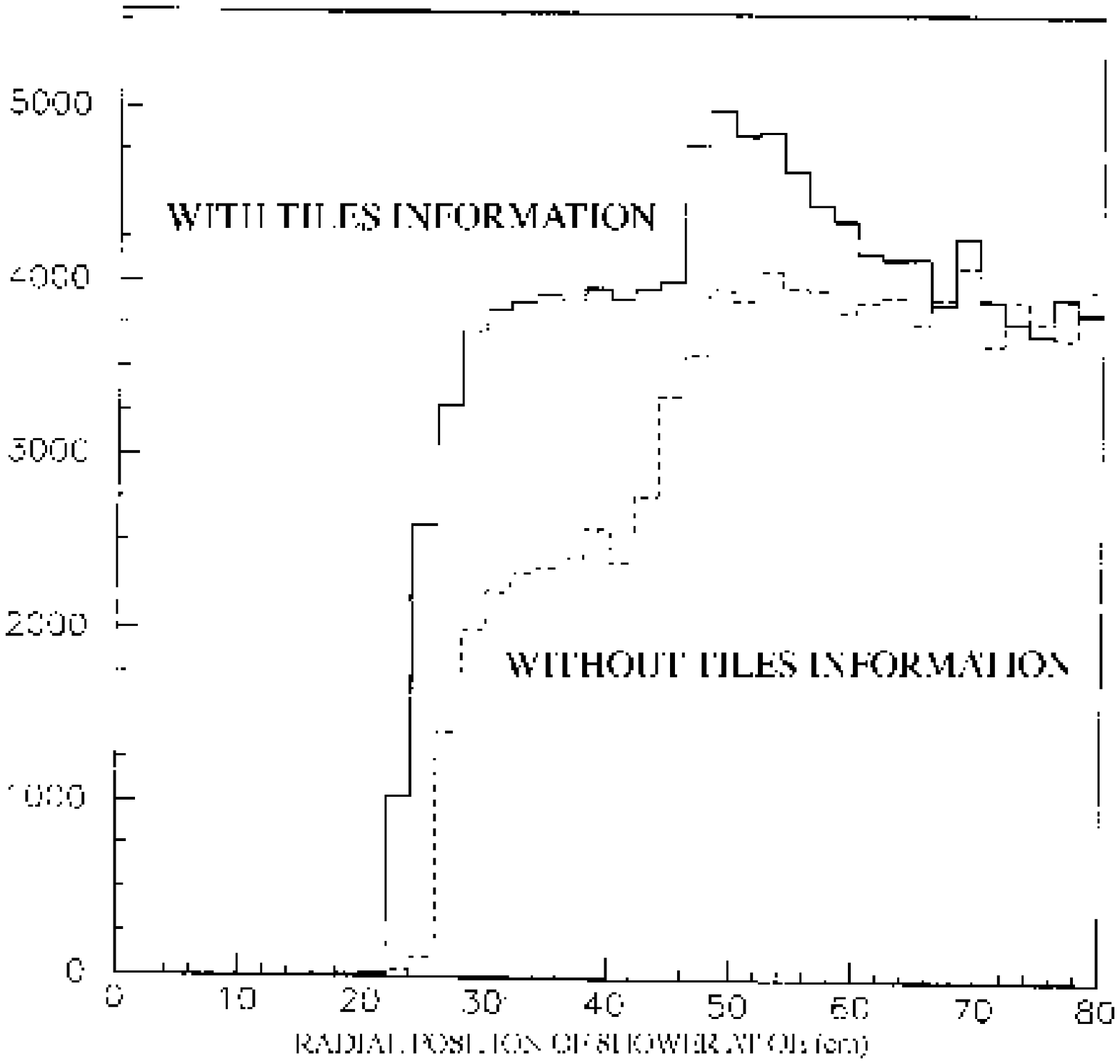,width=5cm,height=5cm}
  \epsfig{file=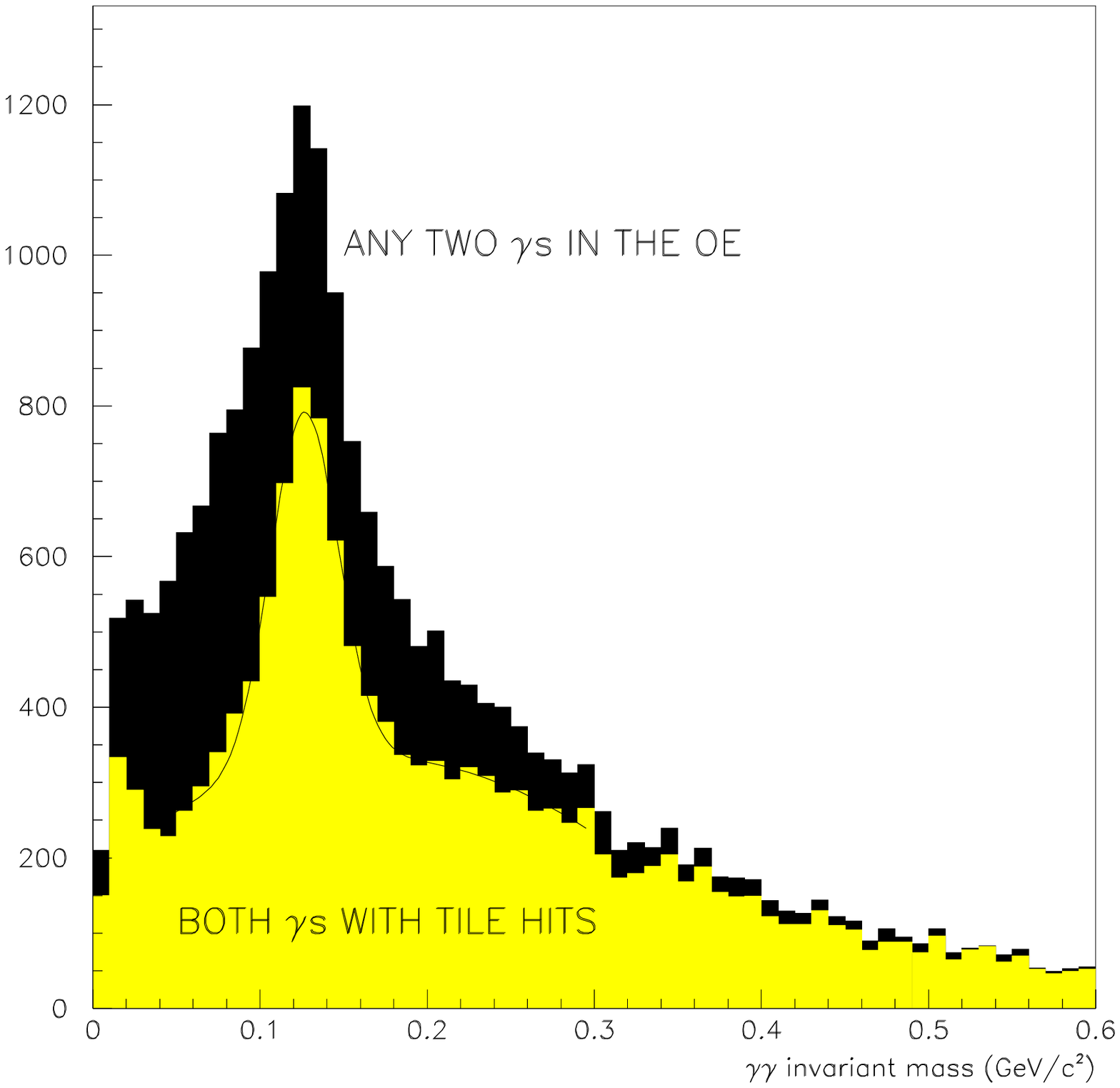,width=5cm,height=5cm}
 \end{center}
 \caption{ (a) Loss of reconstruction efficiency in the small-angle region of
           the Outer em calorimeter; (b) $\gamma\gamma$ invariant mass peak
           with and without tile array hit confirmation.
   \label{fig:bucotile}
 }
\end{figure}
\section{Shower reconstruction strategy}
 The reconstruction algorithm begins 
 with the identification of clustered energy deposits. Clusters of energy
 deposits associated with the projection of charged tracks 
 reconstructed in the magnetic spectrometer are tagged. The reconstruction of
 neutral showers uses the remaining clusters. 
 Energy deposits (in ADC counts units) for each counter 
 are multiplied by the mip equalization constants, and then summed up to
 determine the detected 
 energy  associated to each cluster.
\par
 Pairs of clusters
 in the two orthogonal X-Y views of each OE segment are formed using 
 the energy balance as a criterium, after proper weighing for light
 attenuation  inside the strip counters. 
 Neutral showers are formed by aligned X-Y pairs
 in the different OE segments. The diagonal counters and the tiles are used
 to resolve ambiguities. The use of tile array information improves
 the efficiency of X-Y matching in the small angle region
 (Fig.\ref{fig:bucotile}a) by reducing the number of fake matches
   (Fig.\ref{fig:bucotile}b).
\par
 The coordinates of the shower centroids are determined considering the energy 
 deposited in each counter of the X-Y pairs.
 Once corrected for systematic effects (\S~6), the shower centroids determine
 the   photon incidence point. The $\pi^0$ invariant mass is computed
 by the measurement of relative angle and energy of the decay photons.
 When computing the invariant mass of higher states with one or more
 $\pi^0$ in the final state, the $\pi^0$ invariant mass is fixed at its
 nominal rest value, and the photon momenta are rearranged by means of a 1-C
 fit. The classical algorithm in 
 Ref.\cite{Nakano:1985pi}
 was modified in order to
 take into account space resolution\cite{Gianini98}.
%
%
\par
 Neutral showers are identified as em or hadronic by means of the
 Discriminant Analysis algorithm (\S~7). 
 Finally, the sum of energy clusters longitudinally forming a 
 reconstructed photon track gives the detected
 energy $E_{detected}$. 
 The reconstruction efficiency for single isolated showers was measured 
 using primary Bethe-Heitler $\epem$ pairs. The $\epem$ tracks found in the
 proportional chambers were projected onto the OE front. Shower reconstruction
 was then performed using all the available clusters. The efficiency for 
 reconstructing the shower associated with the electron or positron  track
 was greater  than 95\% over the range $2-20\,{\rm GeV}$, and better than
 90\% over the  range $0.5-2\,{\rm GeV}$.
\section{Energy calibration, linearity, energy and space resolution}
 The response of the calorimeter to photon- and electron-initiated e.m.
 showers, the 
 scale factor $\alpha$ between detected energy and incident energy, and
 the energy   resolution have been studied using a GEANT simulation.
  Simulation predictions have been verified with 
 $\epem$ pairs and electron beam in calibration runs, $\epem$ pairs from the
 process  $\pi^0 \rarr \gamma\gamma, \gamma {\cal N} \rarr \epem$ of
 photon conversion in physics events   for
 electron-initiated showers  constants , and  $\pi^0$ peaks for
 photon-initiated showers constants and    absolute calibration of the
 energy scale. 
\par
\begin{figure}[h]
 \begin{center}
  \epsfig{file=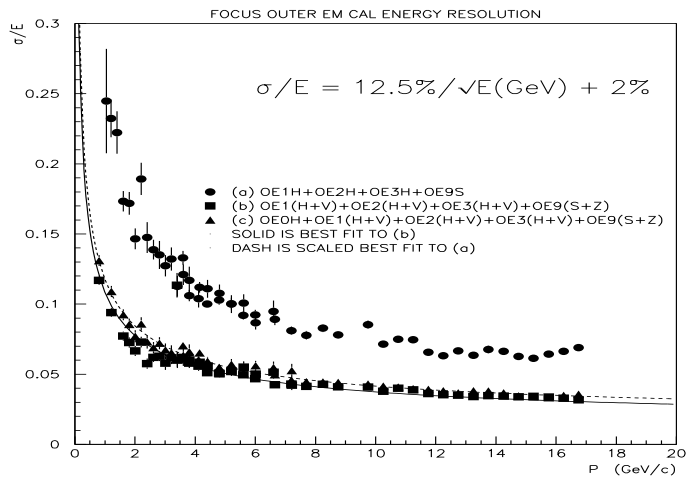,width=10cm,height=7cm}
  \epsfig{file=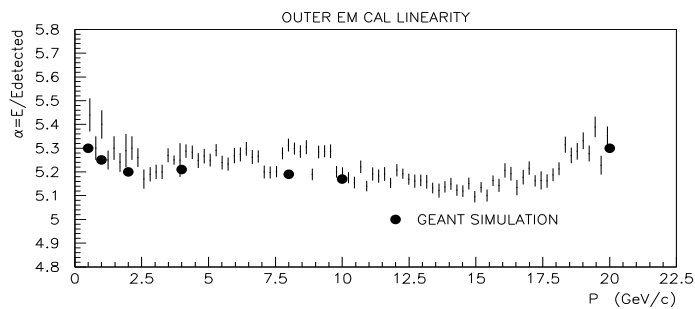,width=10cm,height=5cm}
 \end{center}
 \caption{ Energy resolution (with the constant term added in quadrature)
 and linearity as measured by calibration 
 $\epem$ pairs.
   \label{fig:riso}
 }
\end{figure}

 The detected energy is parametrized as
 $  E_{detected} =  E/\alpha$
 where $E$ is the particle incident energy and $E_{detected}$ is the
 particle energy deposited in the calorimeter active layers.
 Energy linearity and resolution are shown in Fig.\ref{fig:riso},
 in agreement with the simulation predictions. 
\par
%
%
\par
\begin{figure}[t]
 \begin{center}
 \vspace{10cm}
  \includegraphics{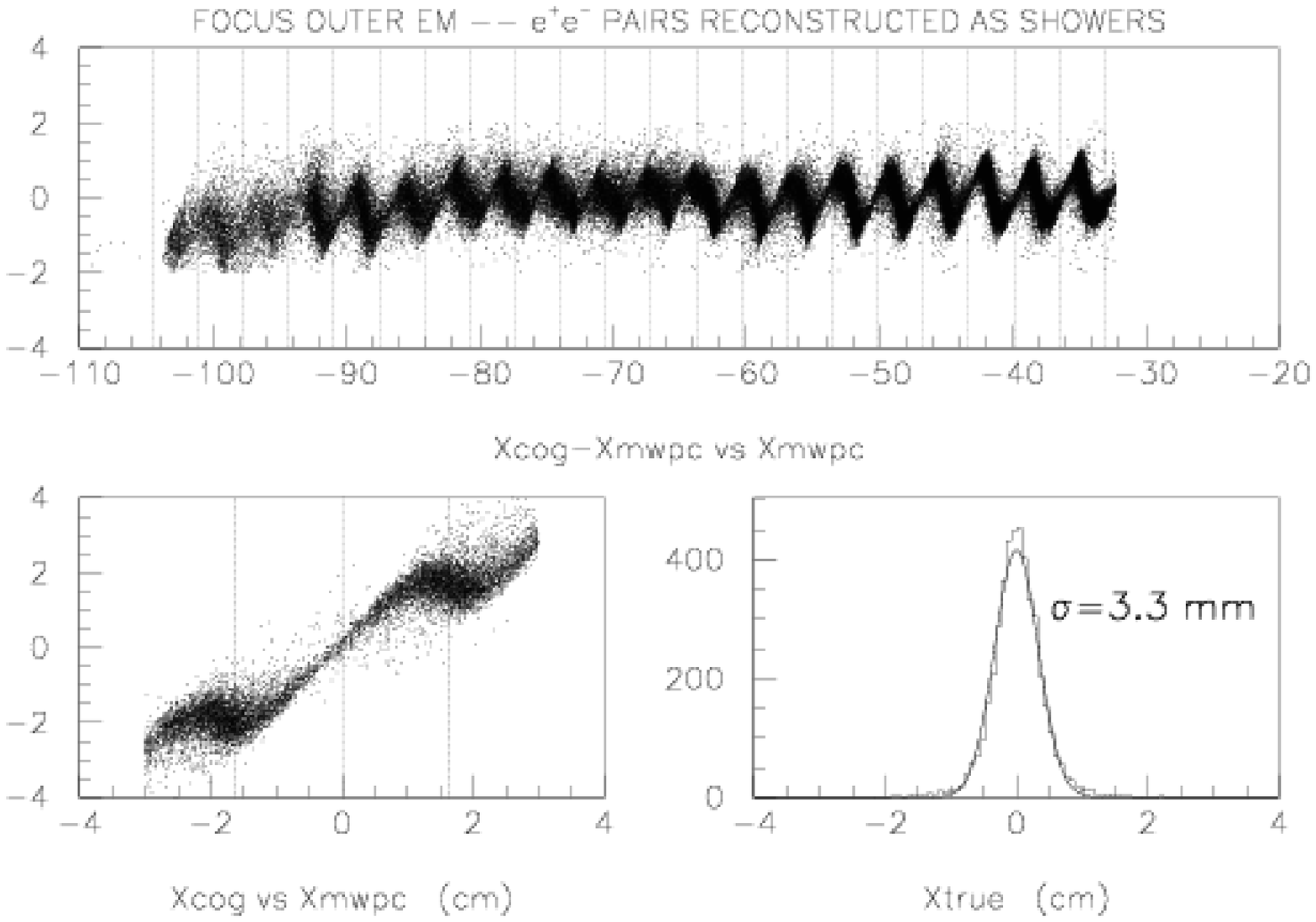}
 \end{center}
 \caption{ a,b) Shower center-of-gravity $X_{cog}$ as measured for a beam of
 calibration electrons sweeping the Outer em at increasing angles. The
 calorimeter response degrades when the em shower reaches the lateral
 boundary of the detector acceptance, thus losing full lateral containment.
 c) Distribution of residuals after correction.
   \label{fig:spazio}
 }
\end{figure}

 The photon impact point $X_{true}$ is determined from 
 the em shower center-of-gravity $X_{cog}\equiv 2\Delta \sum{i} X_i
 A_i/\sum{i}  A_i$, where  $2\Delta=3.3\,{\rm cm}$  is the counter width,
 after applying the  standard correction\cite{Akopdjanov77}
     $X_{true}= b \,{\rm arcsinh }
       \left ( \frac{X_{cog}}{\Delta} {\rm sinh}\frac{\Delta}{b} \right)$
%
 After correction, we  determine the space resolution as from $\epem$
 calibration events  (Fig.\ref{fig:spazio}) to be
 $\sigma(X_{true}) = \pm 0.3\,{\rm cm}$ in the energy range $(1\leq E_\gamma
 \leq 20)\,{\rm GeV}$.
\section{Particle ID}
\begin{figure}[h]
 \begin{center}
  \epsfig{file=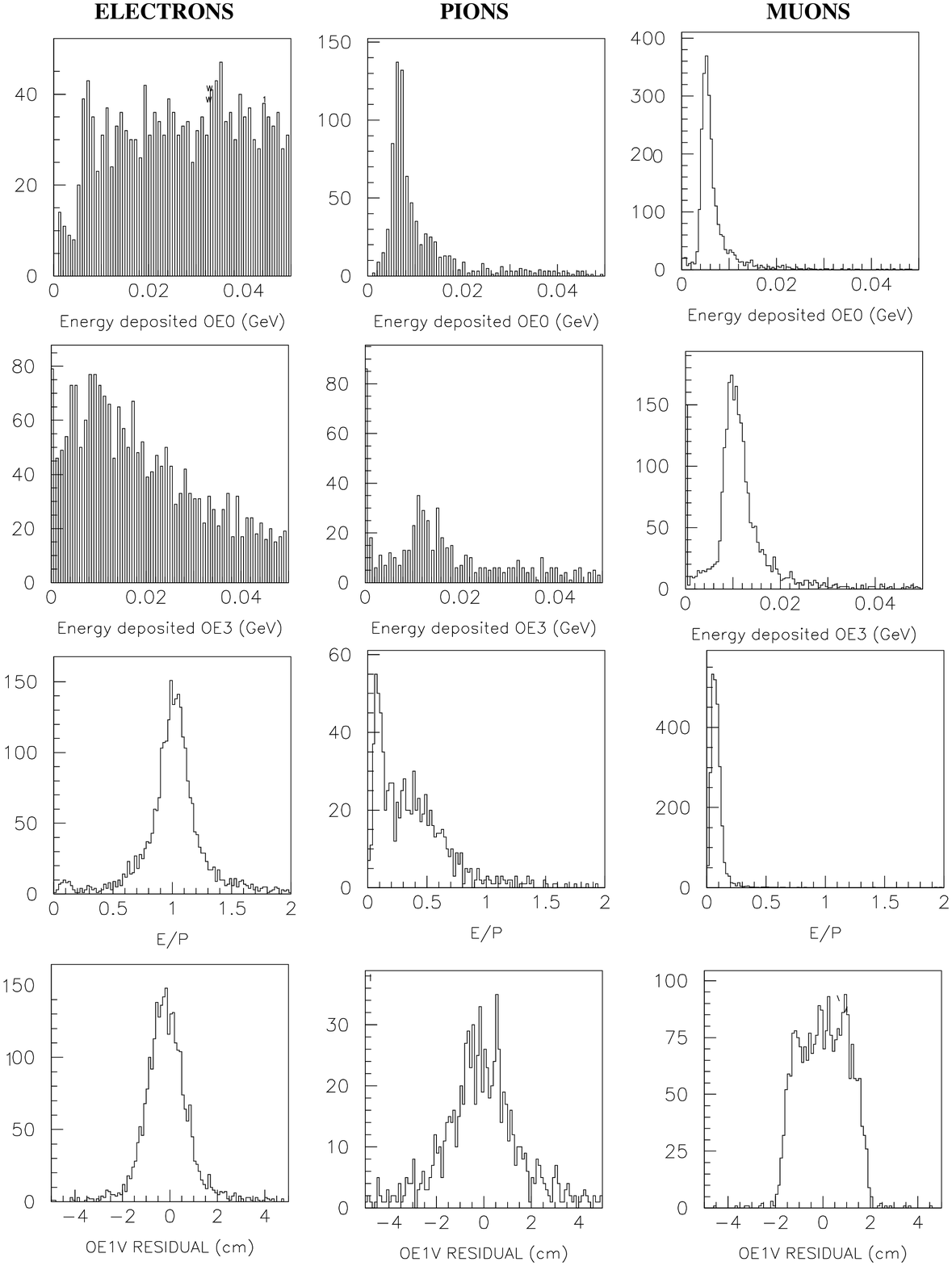,width=15cm,height=20cm}
 \end{center}
 \caption{ Some Discriminant Variables for $e$, $\pi$ and $\mu$.
   \label{fig:pielmu}
 }
\end{figure}

\begin{figure}[h]
 \begin{center}
  \epsfig{file=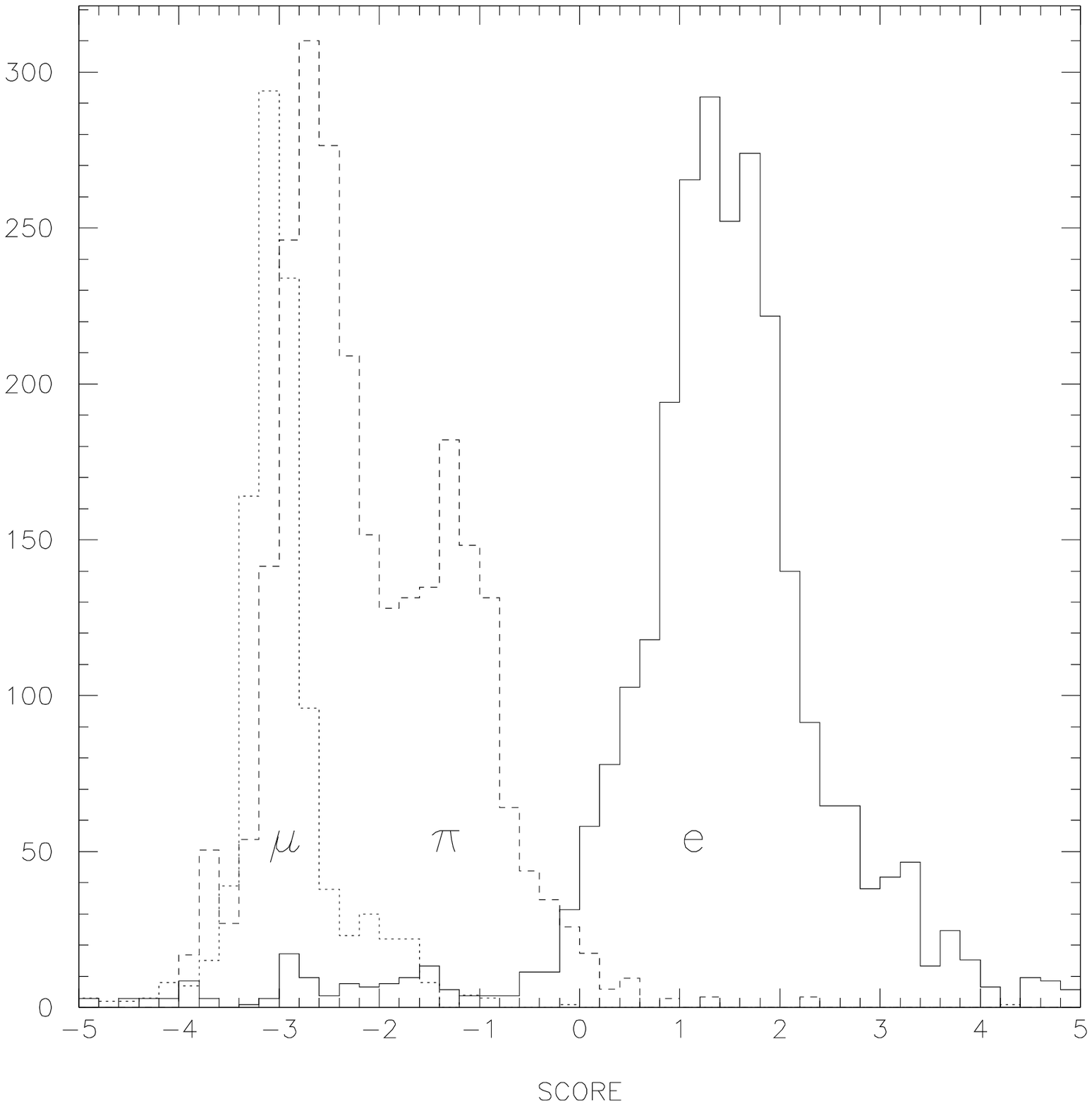,width=15cm,height=10cm}
  \epsfig{file=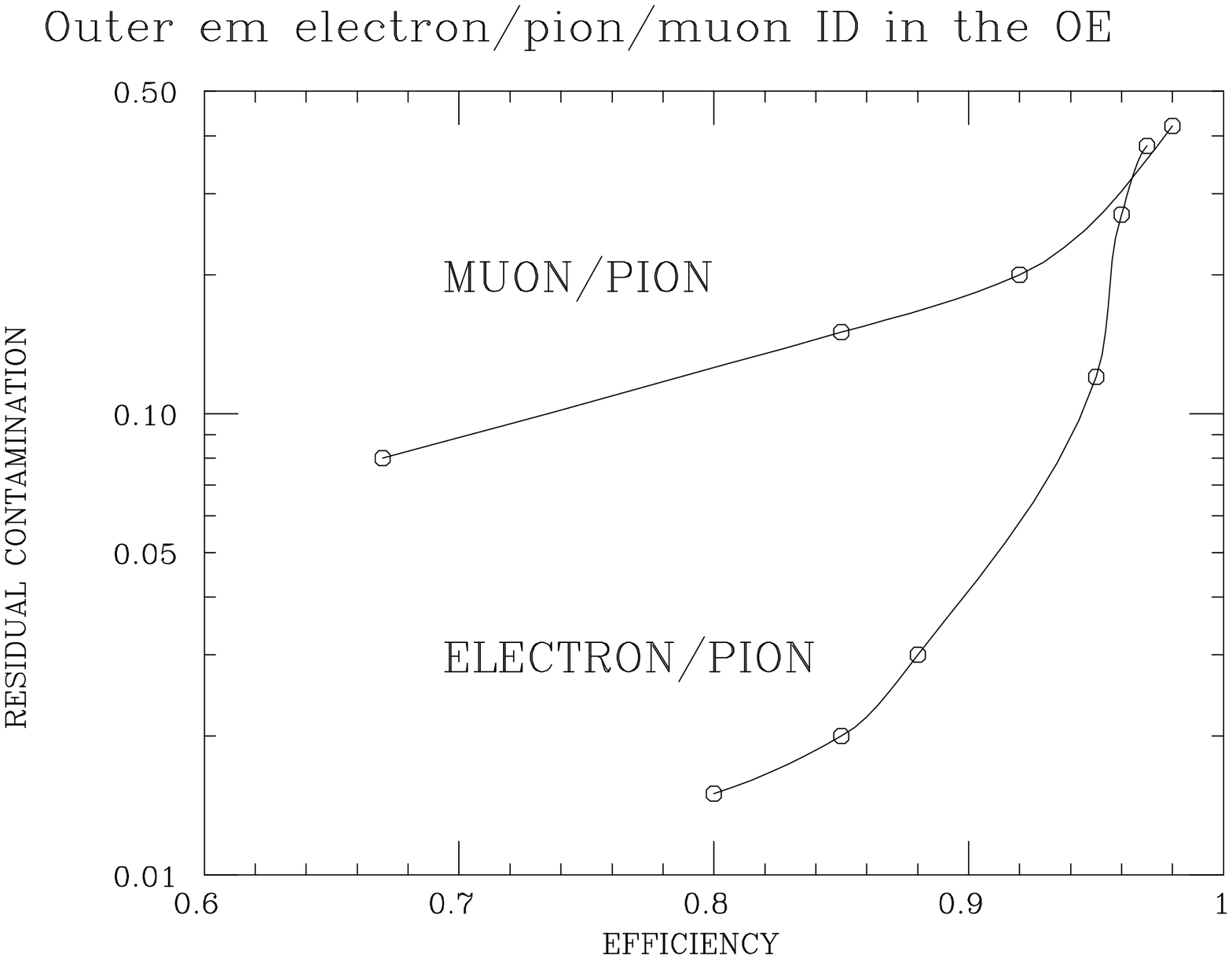,width=10cm,height=15cm,angle=90}  
 \end{center}
 \caption{ a) Distribution of the score variable $S$ for electron, muon, and
 pion training samples; b) efficiency and residual contamination for 
 electron/pion and muon/pion identification.
   \label{fig:da}
 }
\end{figure}

 The Outer em calorimeter extends the $e/\pi$ rejection beyond 
 the Cherenkov momentum range, i.e.,  from 6 to 20 GeV/c.
 The identification algorithms have been developed in the framework of the 
 Discriminant Analysis\cite{spss75}, which allows one to distinguish 
 between  two or more groups of events.
 As first step, we determine a    set of $N$ variables $V_{1,N}$ 
 {\it (Discriminant Variables)}, 
 significantly     different among the M groups of events
$(\{A_j\},\quad j=1,M)$ 
 to be     distinguished (Fig.\ref{fig:pielmu}). 
 A typical set of variables is composed of
 the ratio $E/P$ between the energy measured by the calorimeter and the
 track momentum by the tracking system, the lateral and longitudinal shower
 development  pattern, the cluster centroid residuals and widths.
 Next, the {\it score} function
 $S_A=\sum_{i=1,N} c_i V_i$ is built, and we find coefficients $c_i$ which
 maximize  separation among the           $S_A$, thus   applying one cut on
 $S$ (Fig.\ref{fig:da}a). 
%
 As training samples of known membership we used 
 $\epem$ Bethe-Heitler pairs embedded in hadronic events,
 pions from $K_S \rarr \pi^+ \pi^-$ decays, and  muons from dedicated runs
 with beamdump.
 The overall pion residual contamination obtained for an 
 $85\%$ electron efficiency  is $ 10^{-2}$, while  the pion
 residual contamination is $10^{-1}$ for 85\% muon efficiency
 (Fig.\ref{fig:da}b). 
 Efficiency for muons from $J/\psi$ decay, and rejection of pions compared
 with the 
 Outer muon detector performances are shown in Fig.\ref{fig:phys}~a,b).
\section{Long-term stability}
\begin{figure}[h]
 \begin{center}
  \epsfig{file=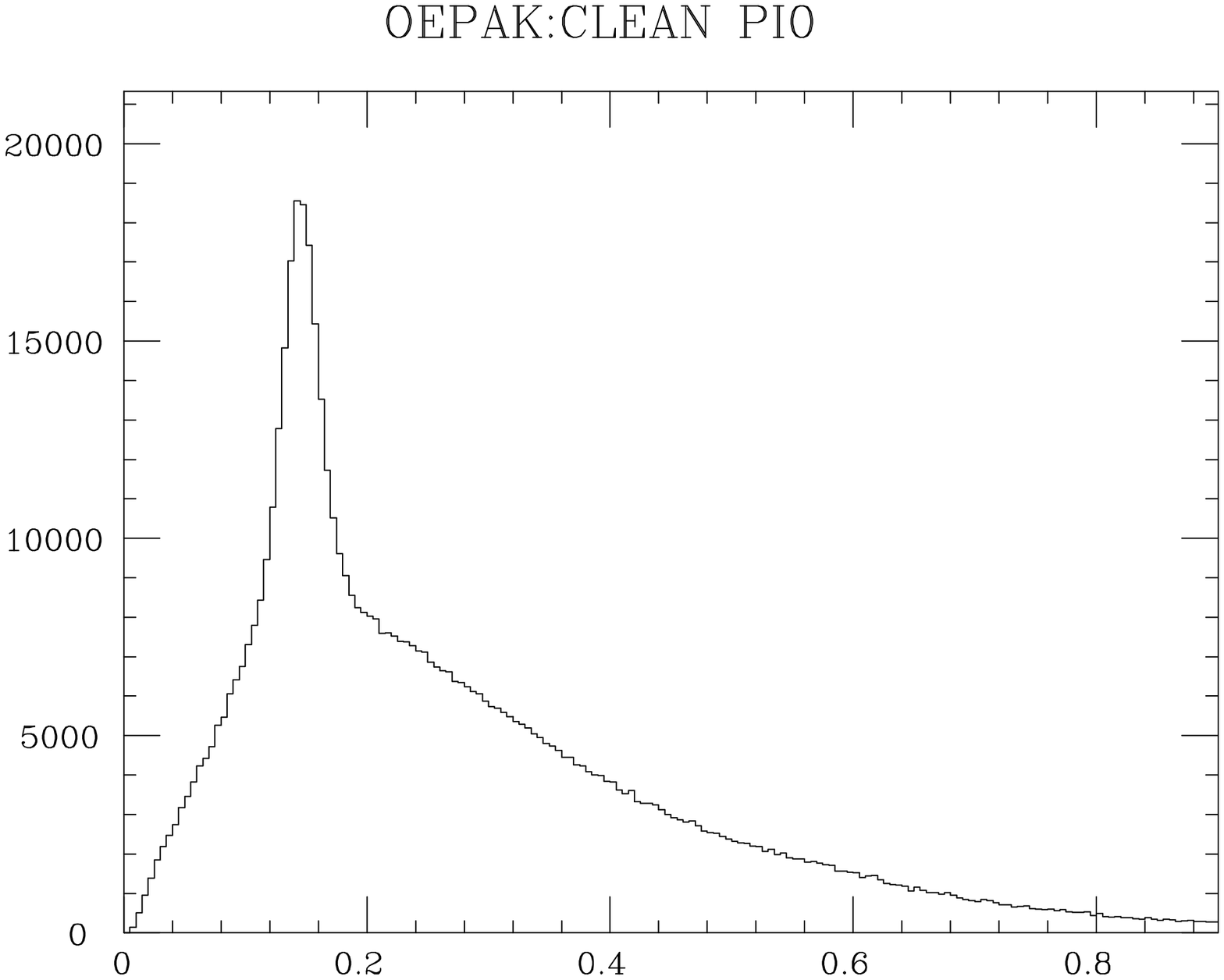,width=5cm,height=4.5cm,angle=90}
  \epsfig{file=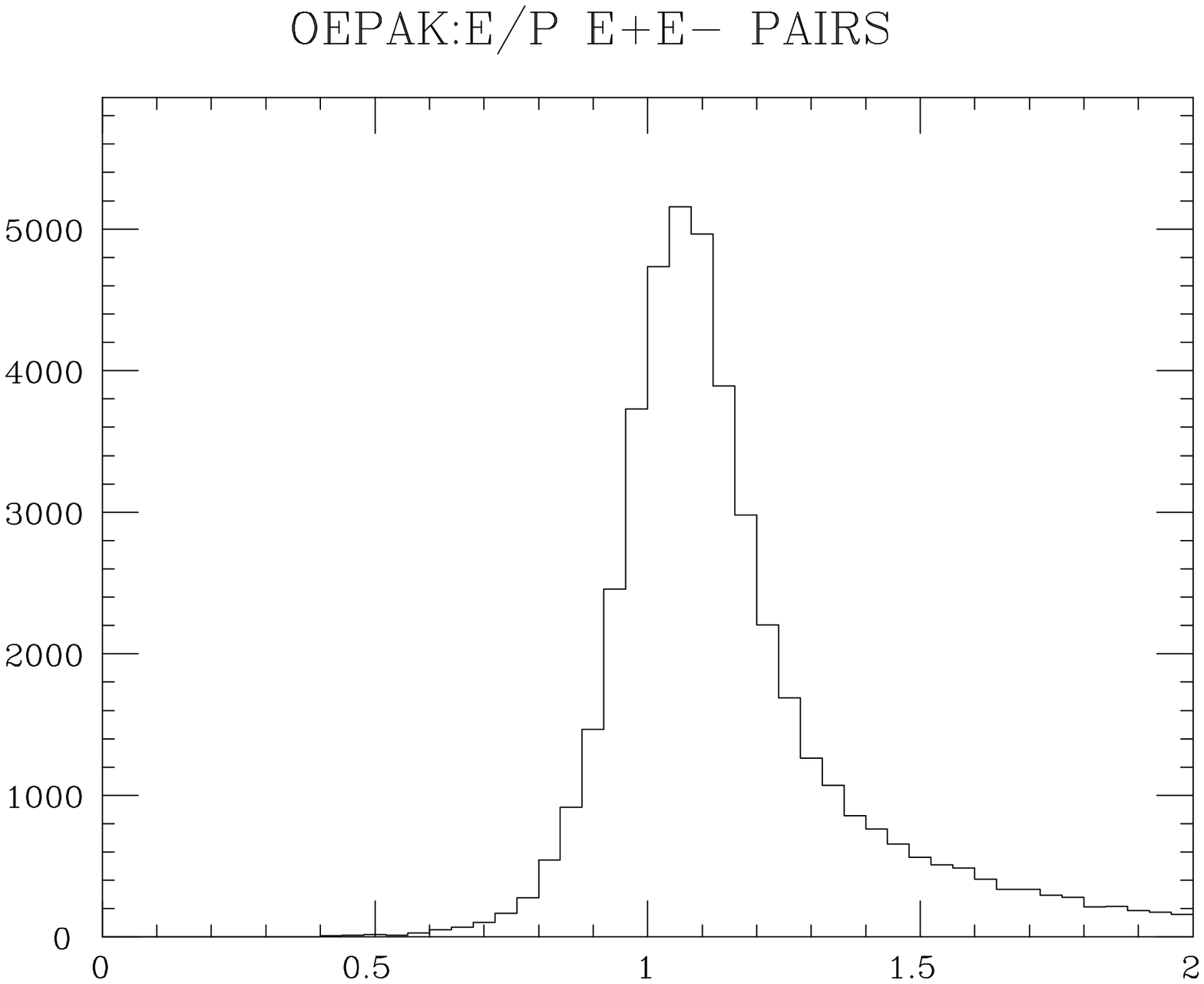,width=5cm,height=4.5cm,angle=90}
  \epsfig{file=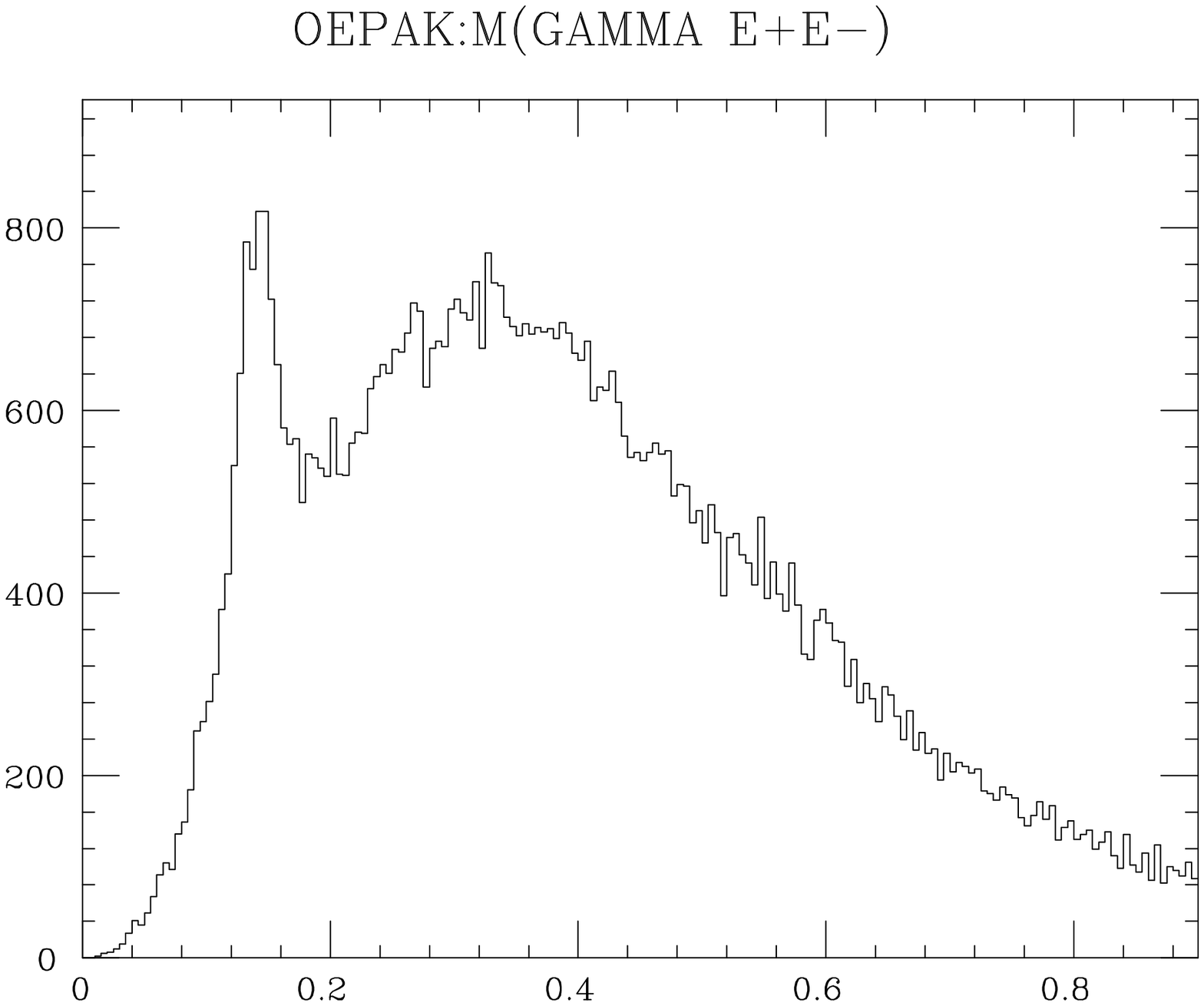,width=5cm,height=4.5cm,angle=90}
  \epsfig{file=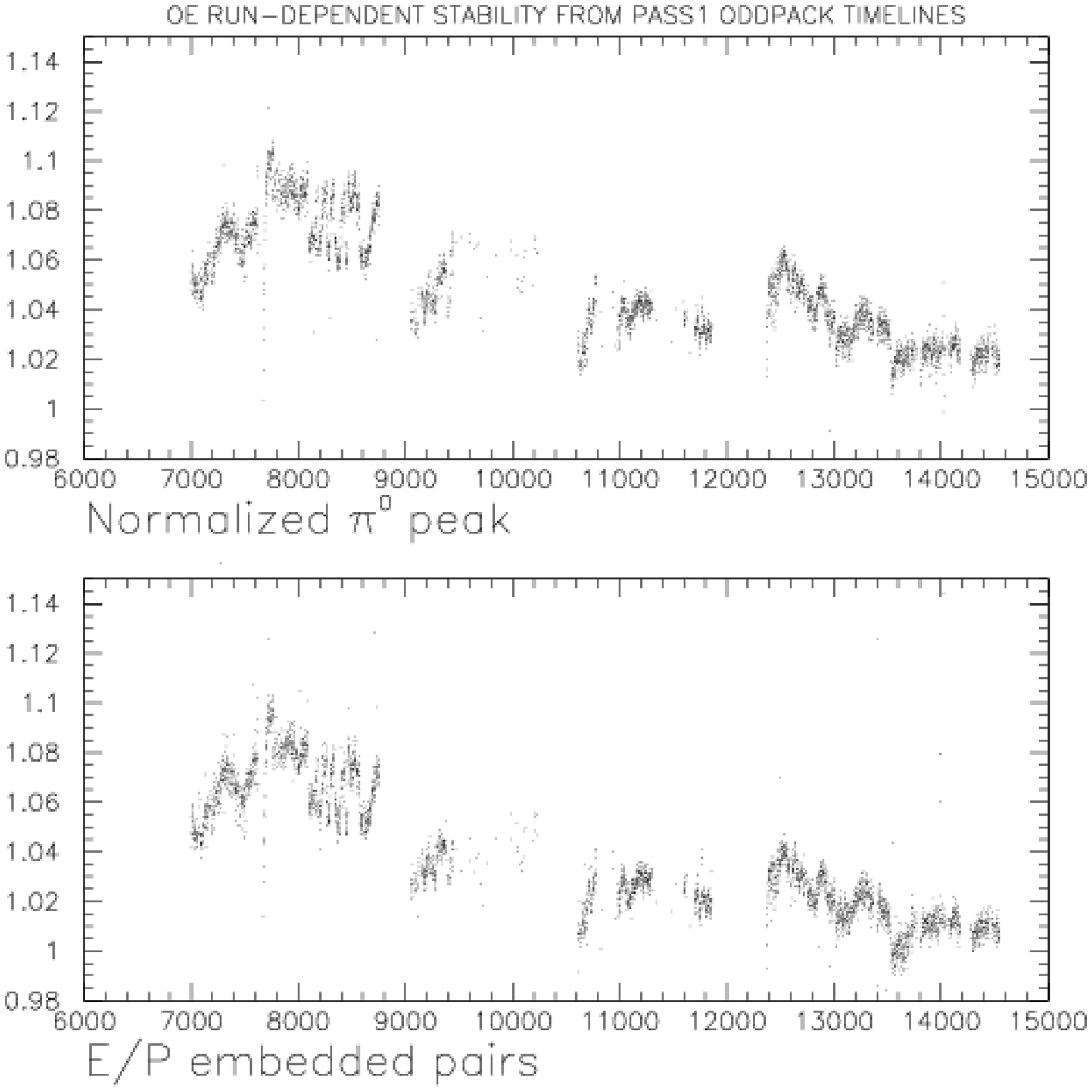,width=15cm,height=8cm}
  \epsfig{file=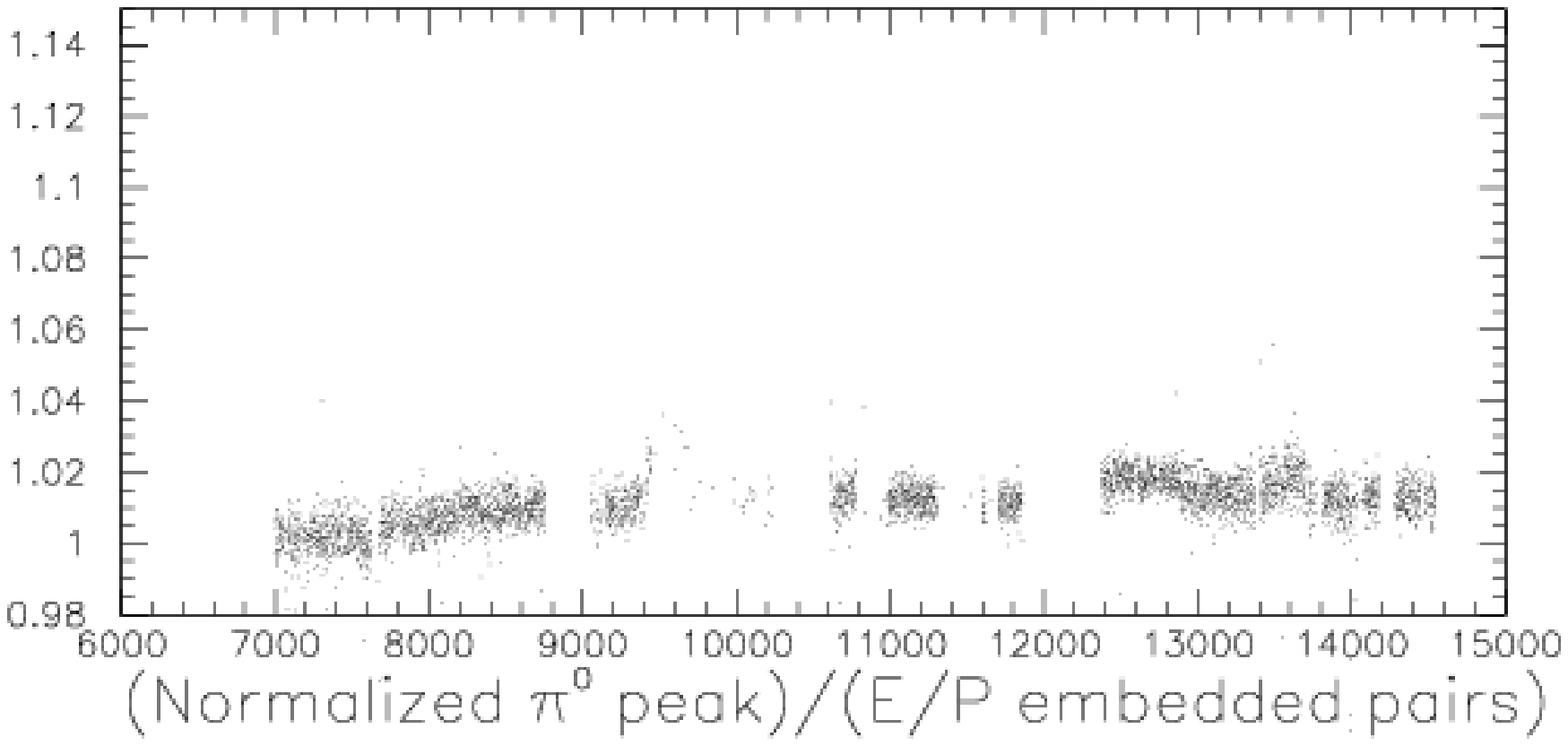,width=15cm,height=4cm}
 \end{center}
 \caption{ Long-term detector stability over 18 months. (top, left to right) 
 $\gamma\gamma$ invariant mass distribution with the $\pi^0$ peak, E/P
 distribution,  and $\gamma \epem$ invariant mass distribution  with the
 $\pi^0$ peak. 
 (bottom) Timelines, as
 a function of  run number (typically one run per hour), of the  E/P and
 $\pi^0$ peak, and  the E/P peak corrected by the $\pi^0$ peak. After
 correction, the detector response is stable within $\pm 1\%$. The residual
 shift is compatible with the stability of the charged track spectrometer.
   \label{fig:stabi}
 }
\end{figure}

 The calorimeter stability is controlled by monitoring the ADC pedestals 
 between spills, the PMT HV supplies, and the single-channel rough response by
 a $N_2$ laser light source\cite{Bianco:1991gu}.
 The availability of a muon beam halo over the entire area of the Outer em
 allows muon calibrations to be performed regularly. The long-term,
 fine-grained 
 run-dependent stability is given by exploiting physical signals in events,
 namely 
\begin{enumerate}
 \item $\pi^0 \rarr \gamma \gamma$ (Fig.\ref{fig:stabi}~a)
 \item E/P for electrons ($\epem$ Bethe-Heitler pairs embedded in 
       hadronic events, Fig.\ref{fig:stabi}~b)
 \item $\pi^0 \rarr \gamma \gamma$,  with one $\gamma$ conversion $\gamma
   {\cal N} \rarr e^+e^-$  (Fig.\ref{fig:stabi}~c)
\end{enumerate}
\par
 Results are summarized in Fig.\ref{fig:stabi}~d,e,f.
 Electron and $\pi^0$ signals in hadronic events (Fig.\ref{fig:stabi}~b,a)
 can track the    shifts in detector response up to a stability of $\pm
 1\%$ over the  entire     18-month data-taking period.
 Work is in  progress to study  process (3) (Fig.\ref{fig:stabi}~c), which will
 provide  on an event-by-event basis the photon energy calibration constant,
 electron and positron E/P, and a check of the  stability of the
 measurement of P, as performed by the magnetic spectrometer. 
\section{Physics results}
\begin{figure}[p]
 \begin{center}
 \vspace{15cm}
  \includegraphics{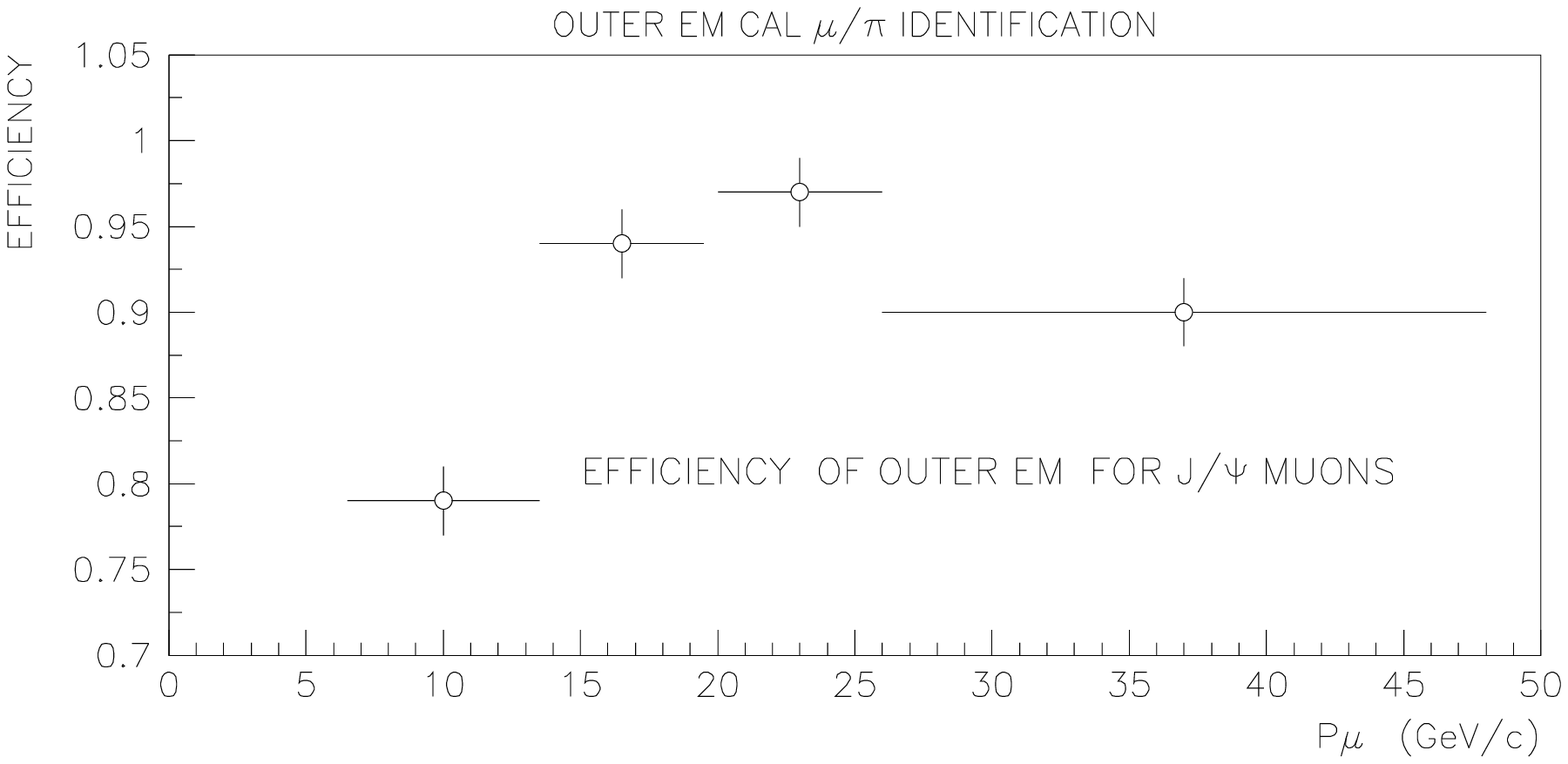}
  \includegraphics{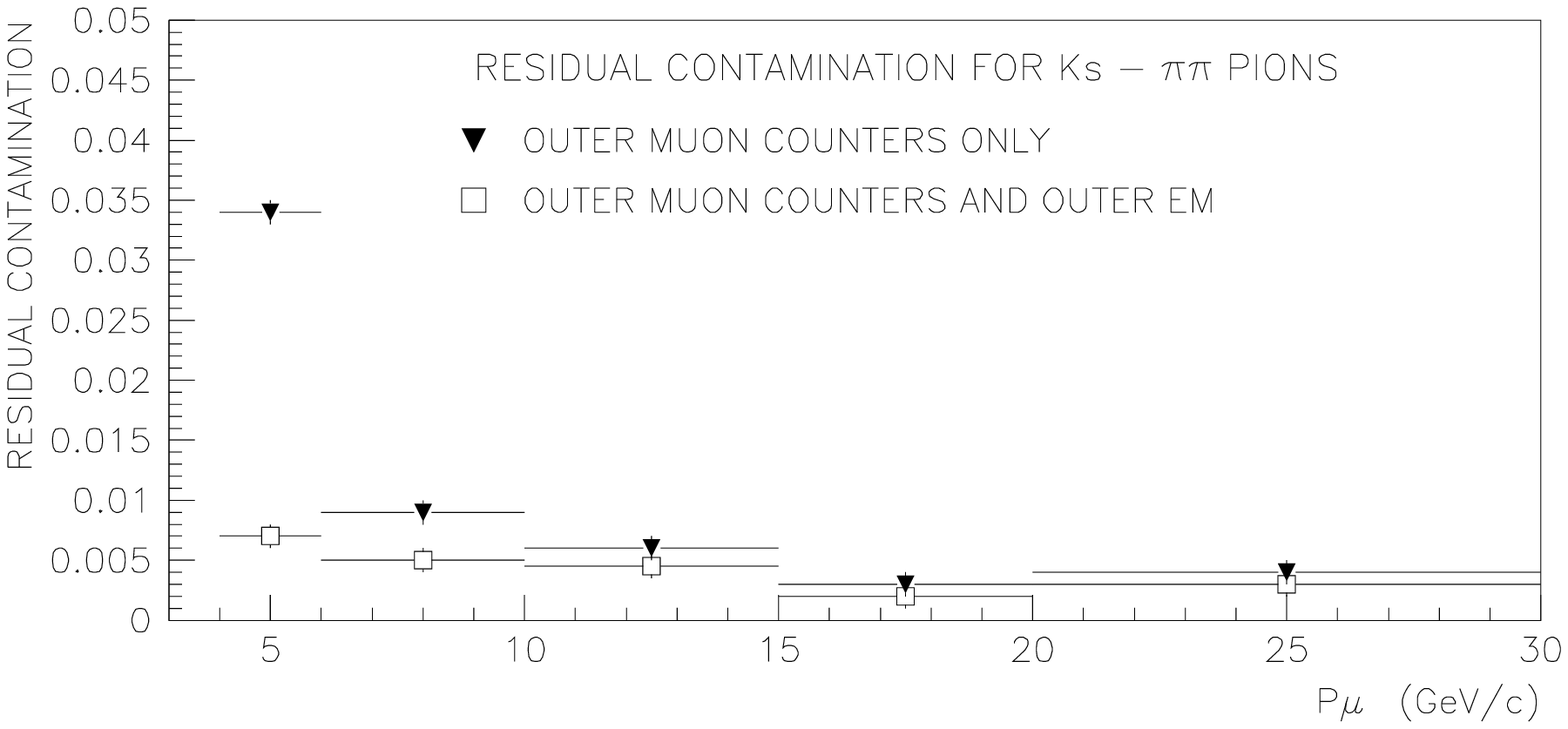}
  \includegraphics{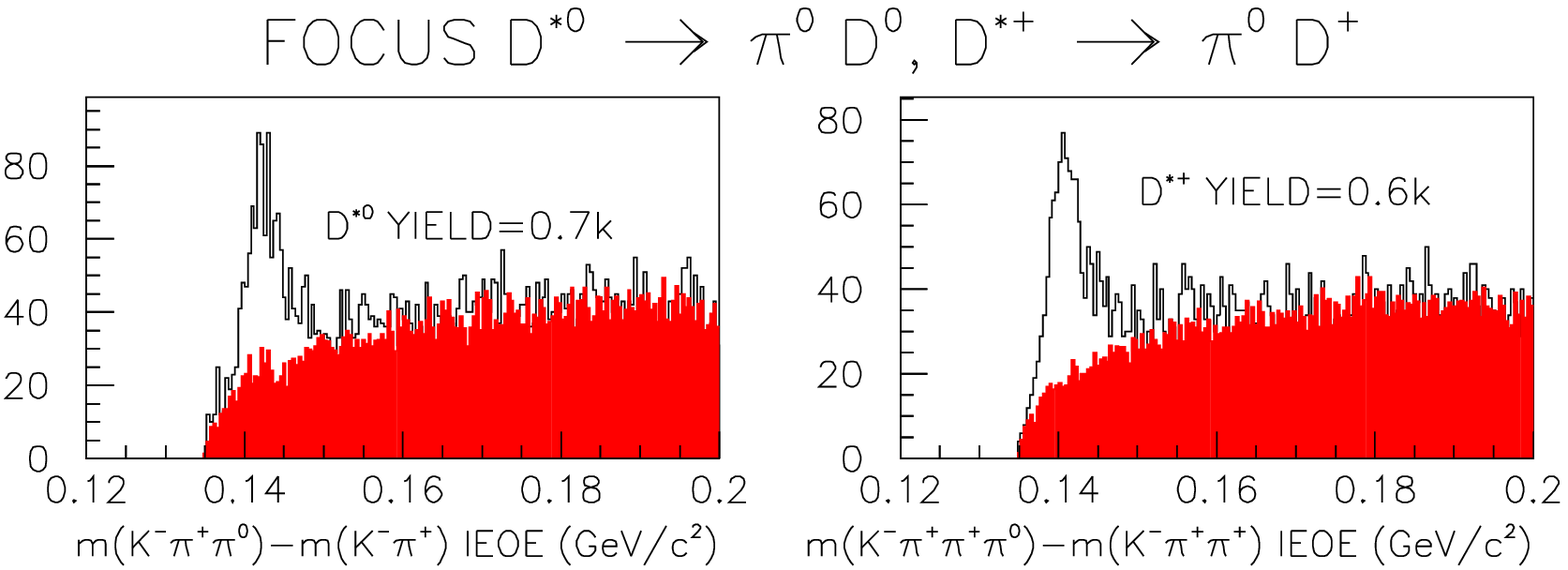}
  \includegraphics{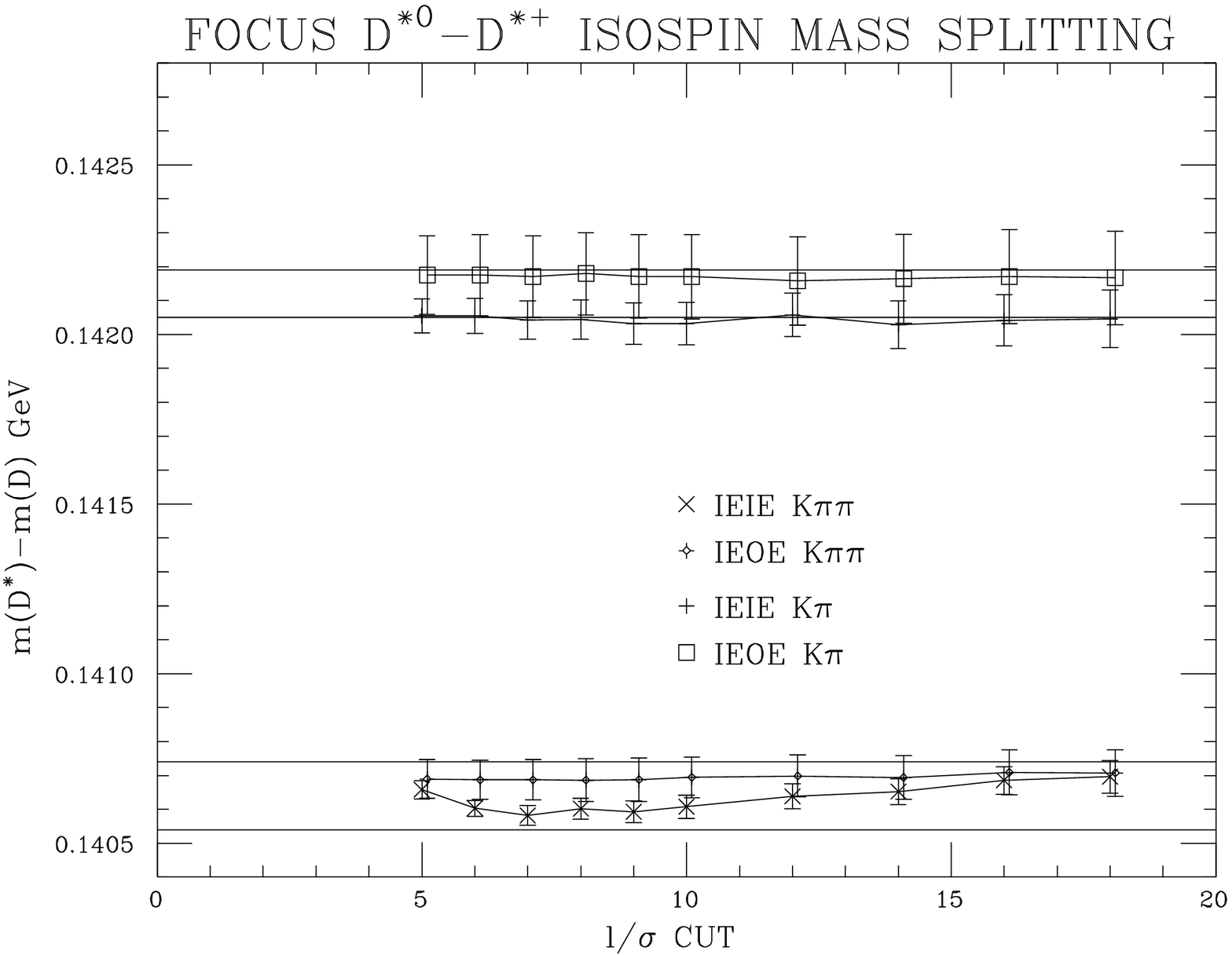}
  \includegraphics{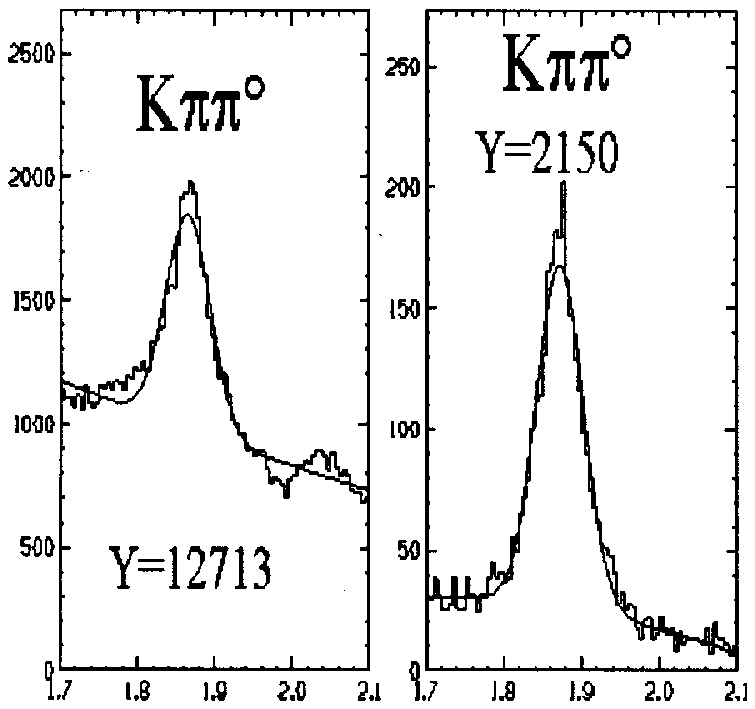}
  \includegraphics{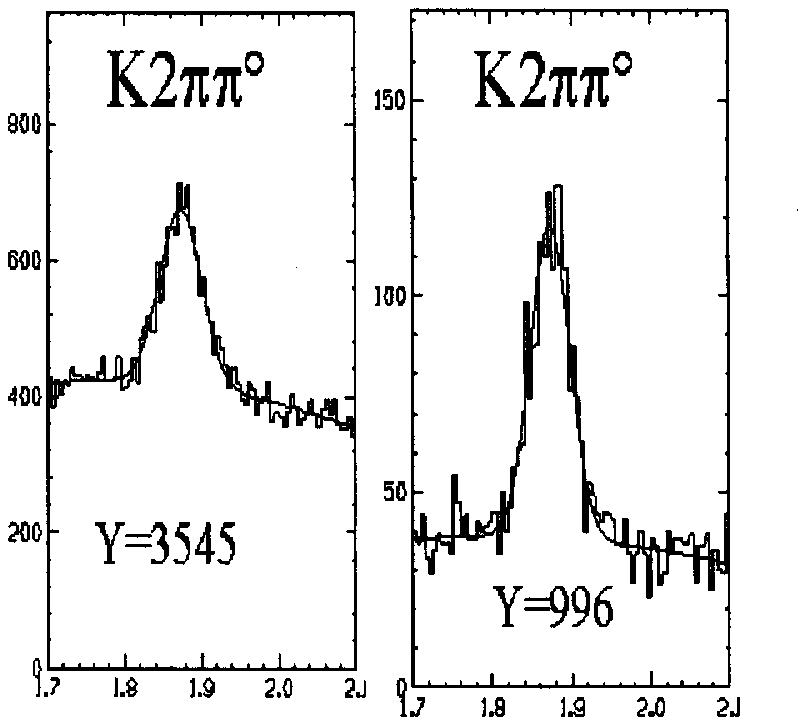}
  \includegraphics{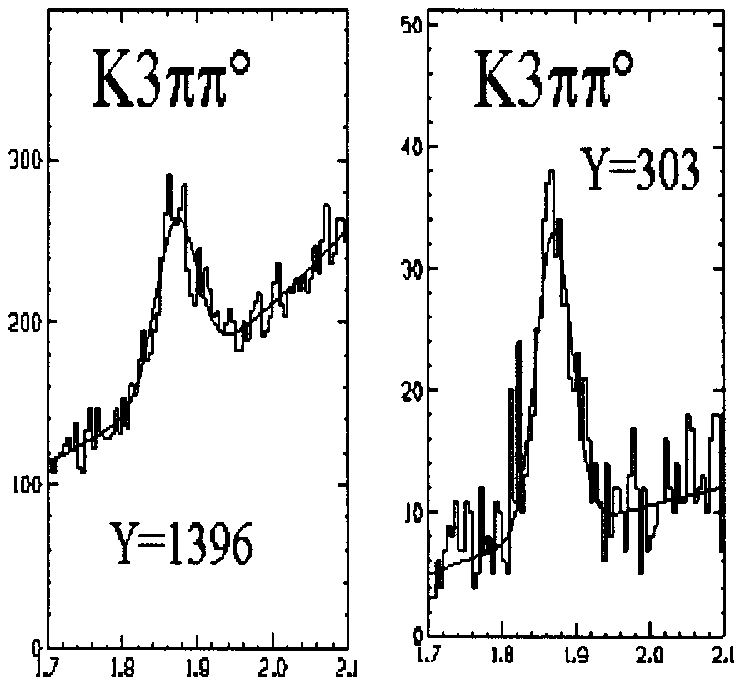}
 \end{center}
  \caption{ FOCUS preliminary results on a fraction of the 1996-97
 dataset. From top to bottom: efficiency on muons from $J/\psi \rarr
 \mu^+\mu^-$  and rejection  of pions from $K_s\rarr \pi^+\pi^-$; $D^{*0}$
 and $D^{*+}$ decays with $\pi^0$ in the final state, and comparison of the
 precision attainable on the isospin mass splittings compared to world
 average, as functions of detached vertex significance $\ell/\sigma$;
 several decays of $D^+$ and $D^0$ mesons with $\pi^0$ in the 
 final state, with a selection of preliminary background-reducing cuts.
   \label{fig:phys}
 }
\end{figure}

   Clean signals of the decays of charm mesons and baryons in many channels
  have been obtained by using the Outer em electron
 and muon identification and the $\pi^0$ reconstruction  (Fig.\ref{fig:phys}). 
\par
 The geometrical acceptance of the Outer em for electrons and muons from
 semileptonic charm decays is about 30\%, extending especially in the
 low-momentum region $P\sim {\rm 4\, GeV/c}$.  
  The detection of charm semileptonic decays with  the electron momentum
 in the acceptance of the Outer em is of  
 particular interest in view of the possibility of
 determining the $q^2$ dependence for the formfactors (see, eg,
 Ref.\cite{Bianco:1999tv}  for an updated review). 
 Studies performed in E687 include the Cabibbo-suppressed
 semileptonic decay $D^0\rarr \pi^- e^+ \nu_e$ (Ref.\cite{Frabetti:1996jj}).
 The efficiency of $\mu/\pi$ rejection was measured on
 $J/\psi$ decays where the muons were identified by the Outer muon counters,
 and found to be better than  80\% from 5 to 50 GeV/c
 (Fig.\ref{fig:phys}a).
 For the $\mu/\pi$ rejection, similarly to the case of $e/\pi$, the
 contribution 
 of the Outer em is  effective especially in the low momentum region
 (Fig.\ref{fig:phys}b),  as  measured on pions from $K_s$ decays.
\par
 Numerous charm decays have been found with $\pi^0$'s reconstructed by the
 Outer  em (Fig.\ref{fig:phys}c-m).
 Thanks to the large statistics and low level of background, precision
 measurements such as the isospin mass splittings $m(D^{*+})-m(D^{*0})$ of
 excited charm meson 
 states  have been initiated. Preliminary results\cite{Wiss99}
 (Fig.\ref{fig:phys}e) show
 how well the attainable precision compares with the best results obtained 
 by $\epem$ experiments using crystal calorimetry\cite{Bortoletto92}.
\par
  In conclusion, the ten-year operational experience of the Outer em
 calorimeter  of FOCUS
 shows how a detector based on conventional techniques is able to perform
 consistently and provide competitive physics results.
 The implementation 
 of a scintillator tile tiebreaker has increased shower reconstruction
 efficiency in the small-angle region, and has considerably cleaned the
 $\pi^0$ peak of spurious combinations.
 Long-term response stability of 1\% is attained by cross-calibration
 between E/P for electrons and $\pi^0$ peaks in hadronic events, while 
 Discriminant Analysis is used to provide $e/\pi$ and $\mu/\pi$ identification.
\par
\section*{Acknowledgements}
%
 We should like to thank L.~Daniello, P.L.~Frabetti. L.~Perasso,
 D.~Torretta, E.~Meroni and A.~Sala for  help in the early stages 
 of construction and operation of the Outer em calorimeter. We also thank
 J.~Mansour, 
 G.~Boca and G.~Apollinari for help and advice on fiber splicing, and
 R.~Justice, E.~LaVallie
 and K.~Gray for on-the-ground help at Fermilab. Help during
 data taking by F.Vasquez-Carrillo and A.~Sanchez is gratefully
 acknowledged. We should like to thank J.~Wiss,
 M.~Nehring and C.Cawlfield for discussions on $\pi^0$ and semileptonics
physics. Finally, 
 we thank the conference organizers for a completely 
 successful conference and the Proceeding Editors for their patience.


\begin{thebibliography}{99}
\bibitem{Frabetti:1992au}
 P.L.~Frabetti {\it et al.},
 Nucl.\ Instrum.\ and Meth.\ {\bf A320}, 519 (1992).
\bibitem{Bianco:1986gf}
 S.~Bianco {\it et al.},
 Frascati preprint LNF-85-49-R.
\bibitem{bologna82}
 G.~Bologna {\it et al.}, Nucl. Instr. and Meth. {\bf 192} (1982) 315.
\bibitem{Apollinari:1992zf}
 G.~Apollinari, D.~Scepanovic and S.~White,
 Nucl.\ Instrum.\ and Meth.\ {\bf A311}, 520 (1992).
\bibitem{Akopdjanov77}
 G.A.~Akopdjanov {\it et al}, Nucl. Instr. and Meth. {\bf 140}, 441 (1977).
\bibitem{spss75}
 Various Authors,  Statistical Package for the Social Sciences (SPSS),
 McGraw-Hill, New York (1975).
\bibitem{Nakano:1985pi}
 I.~Nakano and K.~Miyake,
 Jap.\ J.\ Appl.\ Phys.\ {\bf 24}, 506 (1985).
\bibitem{Gianini98}
 G.~Gianini, {\it The space variables in the $\pi^0$ energy estimate},
 FOCUS internal report E831-mem-1998/7.
\bibitem{Bianco:1991gu}
 S.~Bianco {\it et al.},
 Nucl.\ Instrum.\ and Meth.\ {\bf A305}, 48 (1991).
\bibitem{Bianco:1999tv}
 S.~Bianco,
 invited review at the XIX Physics in Collision, Ann Arbor (USA), June 1999,
 hep-ex/9911034.
\bibitem{Frabetti:1996jj}
 P.L.~Frabetti {\it et al.}
 Phys.\ Lett.\ {\bf B382}, 312 (1996).
\bibitem{Frabetti:1994di}
 P.L.~Frabetti {\it et al.}
 Phys.\ Lett.\ {\bf B331}, 217 (1994).
\bibitem{Wiss99}
 J.Wiss [for the FOCUS Collaboration], APS Centennial meeting, May 1999,
 Atlanta (USA).
\bibitem{Bortoletto92} D.~Bortoletto et al. PRL {\bf 69} (1992) 2046.
\end{thebibliography}
\end{document}